# Contactless photo-induced carrier density control in nanocrystal MoS$_2$ hybrids

Ilka Kriegel,[a,b,c] Nicholas J. Borys,[b,d] Kehao Zhang,[e,f] Adam W. Jansons,[g] Brandon M. Crockett,[g] Kristopher M. Koskela,[g] Edward S. Barnard,[b] Erika Penzo,[b] James E. Hutchison,[g] Joshua A. Robinson,[e,f] Liberato Manna,[a] P. James Schuck[c]

[a] Department of Nanochemistry, Istituto Italiano di Tecnologia (IIT), via Morego 30, I-16163 Genova, Italy
[b] The Molecular Foundry, Lawrence Berkeley National Laboratory, Berkeley, CA, USA
[c] Department of Mechanical Engineering, Columbia University, New York, NY, USA
[d] Department of Physics, Montana State University, Bozeman, MT, USA
[e] Department of Materials Science and Engineering, Center for 2-Dimensional and Layered Materials, The Pennsylvania State University, University Park, Pennsylvania, 16802, United States
[f] Center for Atomically Thin Multifunctional Coatings (ATOMIC), The Pennsylvania State University, University Park, Pennsylvania, 16802, United States
[g] Materials Science Institute and Department of Chemistry, University of Oregon, Eugene, Oregon 97403, United States

**Abstract:**

The ultrathin nature of two-dimensional monolayer semiconductors yields optoelectronic properties which are highly responsive to changes in free-carrier density, making it imperative to masterfully control their doping levels. We report a new photo-doping scheme that quasi-permanently dopes the monolayer MoS$_2$ to extents competing with electrostatic gating. The photo-doping is achieved by coupling monolayer MoS$_2$ with indium tin oxide nanocrystals that can store multiple electrons per nanocrystal after UV illumination. In the hybrid structure, the photo-generated valence band holes in the nanocrystals are filled by MoS$_2$ electrons, photo-doping the MoS$_2$ with holes. Reductions in carrier density by ~6×10$^{12}$ cm$^{-2}$ are observed, equivalent to the storage of ~40 electrons per nanocrystal. Long-range changes proliferating up to 40 μm away from the localized photodoping result from local bandstructure variations in MoS$_2$. These studies reveal novel all-optical carrier density control in monolayer semiconductors, enabling remote-control of local charge density and innovative energy storage technologies.

**Introduction:**

Two dimensional transition metal dichalcogenides (2D TMDCs) have exceptional electronic, optical, mechanical, chemical and thermal properties.[1] The lack of dangling bonds, high mobility, sub-nanometer confinement of carriers within the plane of the material, and unique band structure make them attractive for a myriad of electronic, optoelectronic, and quantum applications.[2–4] Additionally, their electronic structure and optoelectronic properties can be precisely manipulated by electrostatic stimuli[5], stoichiometric defects[6,7] and vacancies,[8] strain,[9,10] and chemical doping,[11] providing unprecedented tunability of the local response of the material.[5,12] Such local manipulation of the optical and electronic properties can be exploited for novel device architectures, suggesting new design principles to create energy funnels for solar energy conversion[10,13,14] or quantum optics.[15]

Light is an attractive tool for locally modulating optoelectronic properties without contacting the material, particularly for applications where bulky, invasive and opaque metal contacts can negatively perturb the material and reduce its performance.[16] In monolayers and few-layer TMDCs, light-matter interactions and optoelectronic properties are governed by an intricate interplay between tightly bound excitons and free-carriers. When illuminated with light, temporary changes to the optoelectronic properties can be achieved that range from localized modification of the conductance and generation of a photocurrent[17,18] to an excitonic Mott transition[19] to negative photoconductivity[20]. While compelling, these temporary changes rely on photoexcited populations that decay on ultrafast timescales. Recently, building on chemical doping via molecular adsorbates, photoactive molecules, such as azobenzene or graphene quantum dots have been used to inject free carriers into TMDCs when illuminated.[21–23] However, in these studies, only a very limited number of carriers per molecule (~1) can be injected, which restricts the extent of the photogating to the highest achievable molecular coverage as well as adsorbate binding/adsorption stability and kinetics.[11]

In this work, we demonstrate that photo-induced electron extraction from monolayer $MoS_2$ (i.e., p-doping) by photoexcited Sn-doped $In_2O_3$ (ITO) nanocrystals changes the carrier density in the monolayer $MoS_2$ by amounts, over timescales, and over length scales that have not been achieved in other photo-doping modalities. In comparison to other works[24–31], the key innovation here is the use of the ITO nanocrystals, as they have the ability to capacitively store multiple photoexcited electrons in their conduction band once the hole is removed chemically by a hole scavenger.[32–35] Hundreds of carriers can be stored in one single, optimized nanocapacitor of several nanometers in diameter.[32] We engineer a hybrid system composed of monolayer $MoS_2$ covered in a layer of ITO nanocrystals where, in effect, the underlying monolayer $MoS_2$ serves as the hole scavenger. In this configuration, a single ITO nanocrystal can extract >40 electrons from the $MoS_2$ – a charge transfer capacity that is not attainable with molecular adsorbates. We find that the electron-transfer from the $MoS_2$ to the ITO nanocrystals is controllable and only occurs when the nanocrystals are excited with above-gap excitation in the ultraviolet (UV), with exposure time as a tool to control the total amount of charge transfer. Analysis of changes in the $MoS_2$ photoluminescence spectra before and after the photodoping reveals that the local optoelectronic inhomogeneity of the monolayer $MoS_2$ drives the phenomena over macroscopic distances, with spatially-dependent measurements showing electron transport and extraction occurring on length scales that exceed several tens of micrometers. The resulting long-distance charge separation makes the charge-transfer quasi-permanent, persisting over timescales of minutes, and possibly longer, highlighting the possibility of *permanent photo-induced* charge separation. Along with being the first demonstration of all solid state photodoping of metal oxide nanocrystals, this hybrid system presents a facile method to achieve wireless, optically triggered p-type photodoping of (or electron extraction from) 2D TMDC semiconductors over a broad range of carrier concentrations, enabling e.g. remote-control and tuning of local charge density in optoelectronic devices and novel light-driven capacitive energy storage.

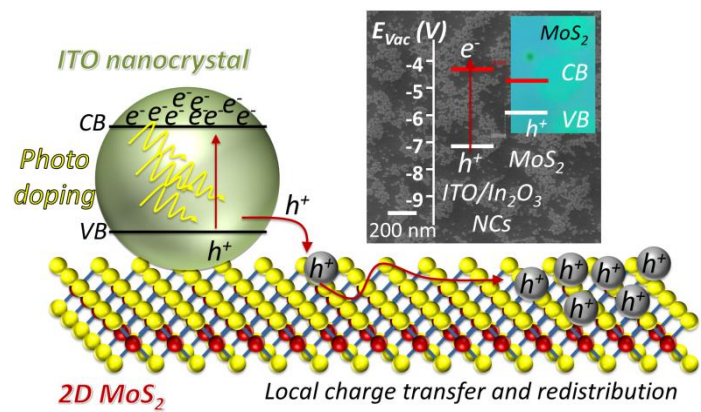

**Figure 1 - Illustration of the hybrid $MoS_2$/ITO nanocrystal structure.** $Sn:In_2O_3$ (ITO)/$In_2O_3$ core/shell nanocrystals (NC) are coupled to monolayer 2D $MoS_2$ (not to scale). Photodoping induces local hole transfer and carrier redistribution in the flake. Inset: Typical SEM image of the ITO nanocrystal film and an optical micrograph of the monolayer $MoS_2$ after deposition of the nanocrystals. After UV light absorption in ITO, an electron is promoted to the conduction band. The hole is removed by transferring to the underlying $MoS_2$ leading to the separation of charges. Together this process can be considered as a photodoping process in which the carrier density in both subunits is quasi-permanently changed.

## Discussion

### Photo-induced electron extraction from MoS$_2$

Fig. 1 summarizes the nanocrystal-TMDC system that we constructed and studied here. The hybrid system, composed of ITO nanocrystals deposited on top of CVD-grown monolayer MoS$_2$ (Fig. 1a), was prepared by spin coating a concentrated solution of the nanocrystals on the monolayer MoS$_2$ crystallites, followed by a subsequent deposition of a thin polymer protection layer (approximately 50-100 nm). The ITO nanocrystals are composed of a Sn-doped In$_2$O$_3$ core with a surface protection layer composed of undoped In$_2$O$_3$[36–38], which serves to passivate surface traps while also aiming to create a favorable heterostructure band alignment that reduces electron-hole recombination and promotes hole transfer by localizing electrons to the core and holes to the surface. Nanocrystal synthesis and characterization are provided in the 'Methods' section. A typical SEM image of the nanocrystals is shown in Fig. 1b together with a representative optical micrograph of the monolayer MoS$_2$ flakes.

Fig. 2 shows the temporal evolution of the photoluminescence (PL) spectrum of the monolayer MoS$_2$ as the hybrid system is illuminated with focused 350 nm UV excitation (see the 'Methods' section). In these measurements (bottom four panels), the UV excitation simultaneously photoexcites the ITO nanocrystals and the underlying monolayer MoS$_2$ allowing us to concurrently monitor how the PL changes as holes are injected into the monolayer MoS$_2$ from the ITO nanocrystals. Prior to exciting the system with the UV excitation, the optical response of an MoS$_2$ flake at a single diffraction limited spot with excitation wavelength of 500 nm (below the bandgap of the nanocrystals) was recorded and is shown in the top panel of Figure 2 (red curve, with emission centered around 1.81 eV). Once the photodoping process is initiated with the UV excitation an immediate blue shift of the PL of the monolayer MoS$_2$ is observed (purple curve, 0 min) progressively shifting to higher energies and intensifying over the course of ~10 minutes (Fig. 2, lower panels; green to yellow curves) until no further changes are observed. Notably, in control measurements without the ITO nanocrystals, the PL spectrum of the monolayer MoS$_2$ does not evolve with time under the same UV illumination (Fig. S1).

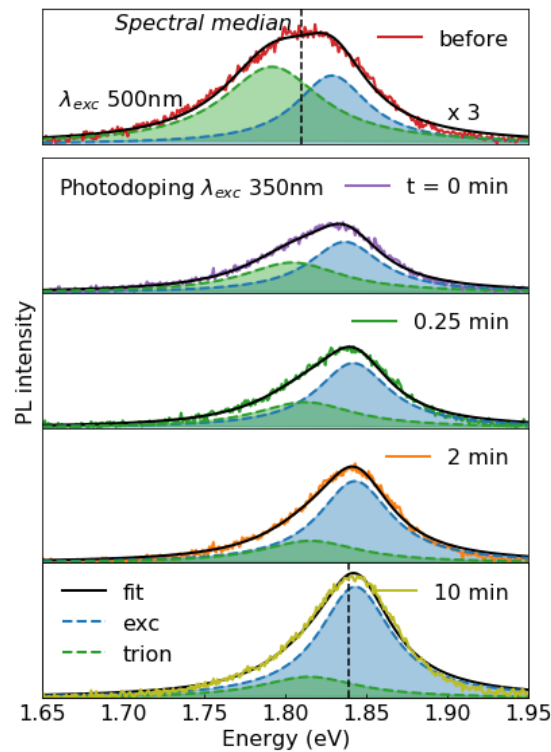

**Figure 2 - Spectral evolution of the MoS$_2$ photoluminescence in the hybrid structure over time during photo-induced carrier extraction.** The initial PL spectrum of MoS$_2$ with excitation wavelength at 500 nm (no ITO excitation) is given in red ($\lambda_{exc}$ = 500 nm). After photodoping with $\lambda_{exc}$ = 350 nm (above the bandgap of the ITO NCs) an immediate blue shift (purple curve) is observed, which further blue shifts and intensifies with increasing exposure up to 10 minutes (compare green to yellow spectra). A fit of the PL (black curve) with two Lorentzian line shapes which are assigned to the exciton (blue curves) and trion (green curves) emission describes the overall PL spectrum well.

Comparing the unphotodoped initial spectrum (top panel, Fig. 2) to the final spectrum (bottom panel, Fig. 2) of the MoS$_2$, it is clear that the PL spectrum not only increases in intensity and shifts in energy, but also changes lineshape. The ultrathin nature of these materials results in strong Coulombic interactions between electrons and holes that yield a rich manifold of excitonic photoexcitations that are stable at room temperature and become apparent in their PL.[2,11,23,39]. In MoS$_2$ and other TMDCs, the most prevalent of these are neutral excitons and charged excitons (or trions). Monolayer MoS$_2$ is typically n-doped,[2,40] so the trions are composed of a neutral exciton bound to an additional electron. The relative spectral weights of the exciton emission and trion emission depend on the doping level of the MoS$_2$, and for the case of the n-doped system here, as more free electrons are removed (i.e., by photoexcited p-doping from the ITO nanocrystals), the relative spectral weight of the trion will decrease. PL intensity also increases as the relative trion contribution decreases, since the quantum yield of trion emission is notably lower than for free excitons.[2,39]

Such deconvolution of room temperature photoluminescence into its contributions from excitons and trions can therefore be exploited to give information on the variation of doping in TMDCs.[2,11,23,39] In our measurements, a fit to the total PL spectrum with two Lorentzian line profiles is used to extract the relative spectral contributions of exciton and trion states to the overall spectrum (blue and green curves Figure 2). As seen in the top-panel, the PL spectrum of the initial undoped monolayer MoS$_2$ can be deconvolved into emission bands from two states: a low-energy trion and high-energy neutral exciton,[2] showing a significant contribution of trion PL, indicative of an initially high density of negatively-charged free carriers in the MoS$_2$.[2,39,41] Also shown in these panels is the spectral median value before and after 10 minutes of photodoping, where the spectral median is defined as the energy that divides the spectrum into equal amounts of high and low-energy emission. This quantity is a convenient descriptor of the data that captures both peak shifts and changes in lineshape and is employed below to help visualize the spatial dependence of the photodoping process and in Fig. S2 to illustrate the time dependent shift of the PL.

**Temporal investigation of carrier extraction**

As can be seen in Fig. 2, the changes in the MoS$_2$ PL when the ITO nanocrystals are illuminated with UV excitation reflect a reduction in the relative spectral weight of the trion state – and a corresponding increase in free exciton population – in the MoS$_2$. The deconvolved dynamics of the spectral weights and energies of the excitons and trions are shown in Fig. 3. Over the photodoping process, the exciton-to-trion ratio ($I_{exc}/I_{trion}$) increases from ~2/3 to ~4.0 (Fig. 3a). Because the photoexcitation density remains constant in these measurements, and signs of sample degradation were not observed, we attribute this significant increase in the exciton-to-trion ratio to a reduction in the density of negatively charged carriers in the MoS$_2$ due to photo-induced electron extraction into the ITO nanocrystals (i.e., photo-induced hole doping of the MoS$_2$, Fig. 3b). The change in the carrier density is also confirmed in shifts in energy of the exciton and trion states ($E_{exc}$ and $E_{trion}$, respectively; Figure 3b i) where both shift to higher energies with dynamics similar to changes in intensity. Furthermore, the energetic difference between the exciton and trion states (the so-called trion binding energy) decreases from ~37 meV to ~28.5 meV (Fig. 3b ii). This final energy is about 10 meV above the anticipated trion binding energy of 18 meV for intrinsically doped MoS$_2$ (i.e., a free carrier density of 0 cm$^{-2}$) that was extracted from measurements on electrostatically gated MoS$_2$.[2] This difference reflects that at the end of the photodoping process shown here, the MoS$_2$ still remains somewhat n-doped (though the total carrier concentration is significantly reduced), which is consistent with the non-zero relative intensity of the trion state. Remarkably, we find that these changes persist throughout the measurement indicating that the photodoping process is quasi-permanent.

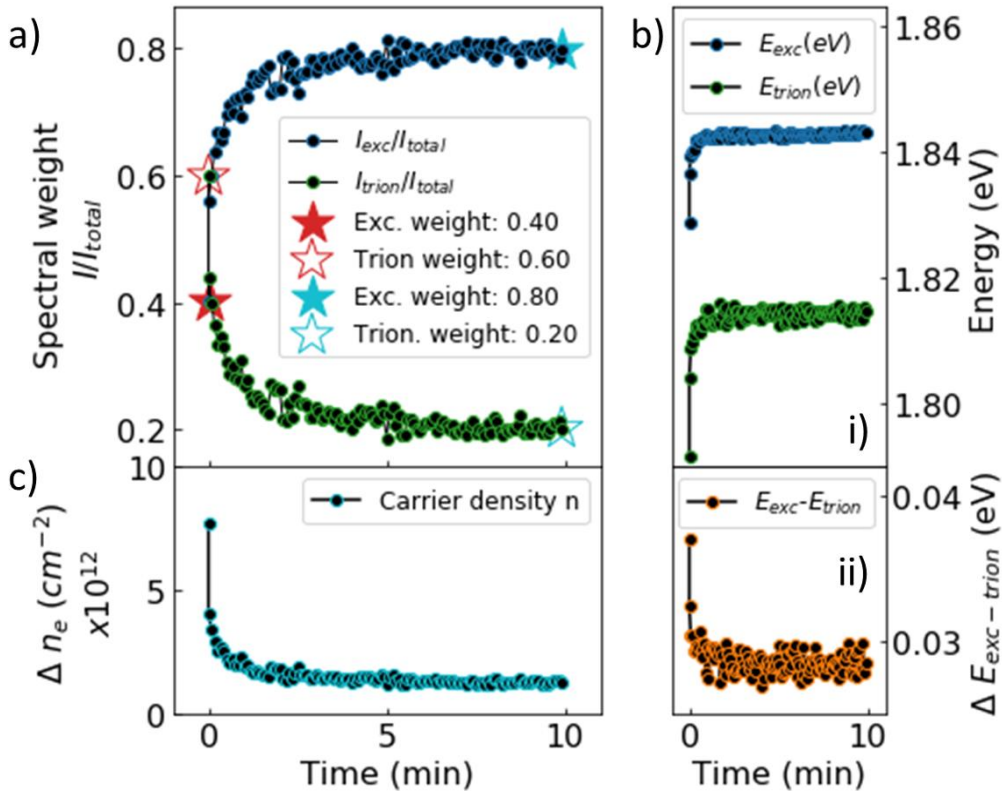

**Figure 3 - Temporal dynamics of photo-induced carrier extraction.** a) Evolution of the spectral weight of the exciton (blue) and trion (green) to the total spectrum over time showing the increased exciton contribution at the expense of the trion upon photodoping. b i) Spectral peak shifts of the separate exciton and trion peaks over time, and ii) energy difference between exciton and trion peak, saturating to ~28.5 meV. c) Time-dependent carrier density extracted from the evaluation of the exciton to trion intensity ratio.

A model can be applied to the changes in the exciton-to-trion ratio to gain a deeper, quantitative insight into the photo-induced changes of the carrier density in the MoS$_2$. For low excitation densities (i.e., when the density of excitons is much smaller than the density of pre-existing carriers), the thermal equilibrium of the free carrier, exciton, and trion populations is given by the Boltzmann distribution.[42,43] The mass action model describing the equilibrium of excitons ($n_x$), trions ($n_{x\text{-}}$) and free carriers ($n_e$) $n_x + n_e \leftrightarrow n_{x\text{-}}$ in the system, was used to describe the variation of the carrier density in quantum well structures[42,43] and adapted to layered 2D materials, such as MoS$_2$ or MoSe$_2$.[11,23,39] Here, we have similarly adapted this model to represent the temporal evolution of the exciton-to-trion ratio, and from that, to extract the variation of the carrier density $n_e$ upon photodoping (Figure 3c):

$$n_e(t) = \frac{n_{x^-}}{n_x}(t) \cdot \frac{4 M_{X_0} m_e}{\pi \hbar^2 M_{x^-}} \cdot k_B T \cdot e^{\left(-\frac{E_B^{x^-}}{k_B T}\right)},$$

where $m_e$ is the effective mass of the electron in MoS$_2$, $M_{X0}$ and $M_{x\text{-}}$ are the effective masses of the exciton and trion respectively, $E_B^{x\text{-}}$ is the intrinsic trion binding energy (from Ref. [2]), and $n_e$, $n_x$ and $n_{x\text{-}}$ denote carrier, neutral exciton and trion densities, respectively. Over the extent of the photodoping process, the carrier density decreases by ~6×10$^{12}$ cm$^{-2}$, which is in the range of carrier density changes achieved in electrostatically gated MoS$_2$.[2,41] Furthermore, the final carrier density reached is approximately 1.5×10$^{12}$ cm$^{-2}$, which again, in comparison to previous works on electrostatically gated MoS$_2$[2], agrees well with the carrier-density anticipated to yield a relative trion intensity of 0.2 with an energetic splitting between the exciton and trion of 28.5 meV.

**Spatial effect of photo-induced carrier extraction**

In many cases, we found that spatially localized photo-induced carrier extraction from the MoS$_2$ into the ITO nanocrystals decreases the carrier density in MoS$_2$ regions that are several microns away from the photoexcited region and thus not directly illuminated. Fig. 4 provides one such example. Prior to the photodoping process, the spatially-dependent PL emission was recorded over the entire extent of the flake (Fig. 4a) using 500 nm excitation as to not trigger the photodoping process and only excite the underlying MoS$_2$. As observed in multiple previous studies,[5,7–9,12,44,45] the PL of the MoS$_2$ is spatially inhomogeneous. For this particular flake, emission from the central region is at a higher energy than the surrounding peripheral edge region. Following this initial PL mapping of the hybrid NC/MoS$_2$ flake, a position in the central region of the MoS$_2$ flake (marked by the star symbol in Figs 4a-d) was illuminated with UV excitation in a focused region of ~1×1 µm$^2$ for 5 minutes to trigger the photodoping process.

Fig. 4b shows the spatially-dependent PL energy of the monolayer MoS$_2$ (again recorded with 500 nm excitation as to not excite the ITO NCs) *after* the localized photodoping at the center of the flake. Notably, the PL in the edge region is shifted to higher energies after the photodoping, with an increased contribution from free excitons. The spatially-dependent before-and-after changes in the MoS$_2$ PL are summarized by the difference plot in Figure 4c. Despite only photoexciting a localized spot in the center region with UV excitation, the most dramatic changes in the MoS$_2$ PL occur in the edge region and reach energetic shifts in the spectral median of up to tens of meV, similar to observations above. Strikingly, the region where the hybrid NC/MoS$_2$ system was illuminated with UV excitation at the center exhibits nearly no change in the PL spectrum. Furthermore, the intensity of the PL in the edge region increases significantly after localized UV exposure (Figure 4d), similar to the effect observed in the temporal variation. A second example of this effect is shown in Figure S 3 in the Supplementary Information. Again, we note that no such changes were observed in the MoS$_2$ PL under UV excitation without the NCs (Figure S 4).

From the spatial dependence of the energy of the PL before photodoping (Fig. 4a), it is clear that the outer regions in these flakes are distinct from the interior. Such distinct edge regions have been observed in MoS$_2$[12] as well as WS$_2$.[7,8] Here, we define a border between an "edge region" and an "interior region" by the contour (black line, Fig. 4a) that traces the midpoint of the range of emission energies (i.e., a spectral median of

1.815 eV in the original PL map). The PL from this edge region is lower in energy before the photodoping process, indicating that the initial carrier density is higher in this region (i.e., there are more trions), which could be due to locally higher strain and/or defects densities, among other causes.[8,12,44] In Figs. 4e and 4f, histograms of exciton and trion emission energy and intensity within the edge and interior regions before and after the photodoping process are presented to provide a statistical summary of the differences (representative examples of the corresponding PL spectra are shown in Figure S 5). In the edge region (Fig. 4e i), the spectral weight of the exciton peak increases while that of the trion decreases, similar to what was observed in Figs 1 and 2. In contrast, no such significant changes are identified in the interior region: the effects of the photodoping are felt most acutely in the edge region, even though the UV excitation was localized in the center of the interior. Further, both the exciton and trion shift to higher energies in the edge region after the photodoping, as statistically summarized in Figure 4f. The shifts in energy are also accompanied by a reduction of the energetic separation between the exciton and trion ($\Delta E_{exc-trion}$) from an average of 29.3 meV to 26.6 meV, again in agreement with changes reported in electrostatically gated monolayer $MoS_2$.[2] These results indicate that the effect of the photo-induced electron extraction from the $MoS_2$ flake is not confined to the UV excitation spot of the laser. Rather, the net change in carrier density occurs preferentially in areas that are locally enriched with trions, which may be indicative of an energetic or electrostatic landscape that promotes a higher local carrier density.

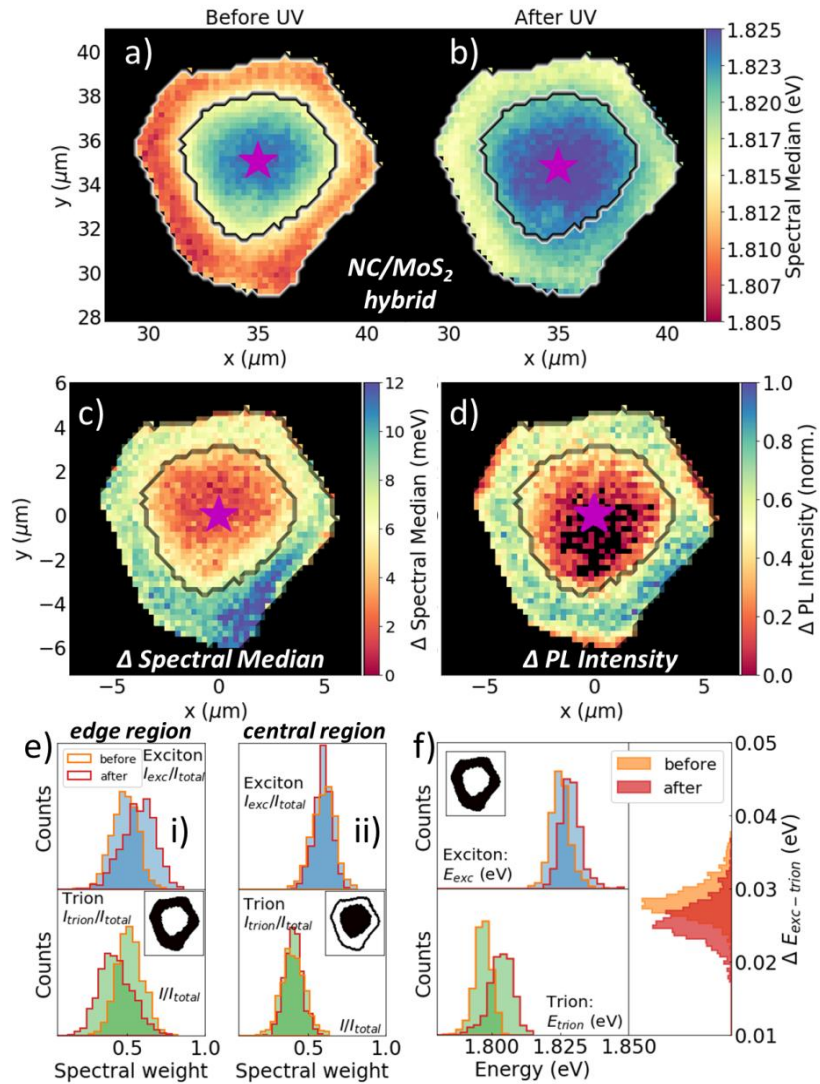

**Figure 4 - Spatial effect of photo-induced carrier extraction in a hybrid NC/MoS₂ flake.** a) Illustration of the overall variation of the PL over an $MoS_2$/ITO NC hybrid flake a) before and b) after exposure to a focused UV laser for 5 minutes at the location marked with a star. Essential differences in the spectral median between the edge and core region of the flake (illustrated by black separating

lines in the plots as guide to the eye) are obvious already before photodoping. c) Plot of the difference of the spectral median of the PL after exposure to UV light illustrating a blue shift after photodoping reaching values up to around 15 meV, particularly obvious in the edge region of the flake, while the core region remains fairly unchanged. d) It is notable that also the intensity of the overall PL in the edge region increases significantly after the localized UV exposure with respect to the core region. e) and f) Statistical summary of the spectral fit of the PL over the entire hybrid NC/MoS$_2$ flake with two Lorentzian line profiles by separately plotting the spectral weight of the exciton (upper panel) and trion (lower panel) before (orange) and after (red) photodoping in the edge (e,i) and core (e,ii) regions. f) Notably for both the exciton and trion, a shift of the center peak was extracted in the edge region as plotted statistically, narrowing the exciton-trion energy separation ($\varDelta E_{exc\ trion}$) from an average of 29.3 meV (red) to 26.6 meV (yellow).

## Large spatial extent of photo-induced electron extraction

To study the spatial extent of the photodoping in the MoS$_2$/hybrid structures, we examine the effect in a larger area of polycrystalline monolayer MoS$_2$ as shown in Figure 5. This spatially extended system is more representative of a continuous film of monolayer MoS$_2$ and is composed of multiple MoS$_2$ flakes that have merged together to form a network of micron-sized crystalline domains separated by grain boundaries. As above, we recorded the spatially dependent PL spectrum of the monolayer MoS$_2$ before and after focused UV excitation to trigger the photodoping process. The difference map of an area that is ~40×60 μm$^2$ (Fig. 5a) shows that after photodoping, the spectral median of the PL increases by up to 30 meV over impressive distances of >30 μm from the region where the photodoping was performed (blue star symbol in Fig. 5a). Notably, the most dramatic increases in the spectral median are localized along the grain boundaries between the individual MoS$_2$ flakes. To quantify the spatial extent, a statistical representation (normalized counts) of the separation from the excitation spot is shown in Figure 5b. At the extreme, changes in the spectral median on the order of 10 meV are detected more than 40 μm away from where the photodoping was performed.

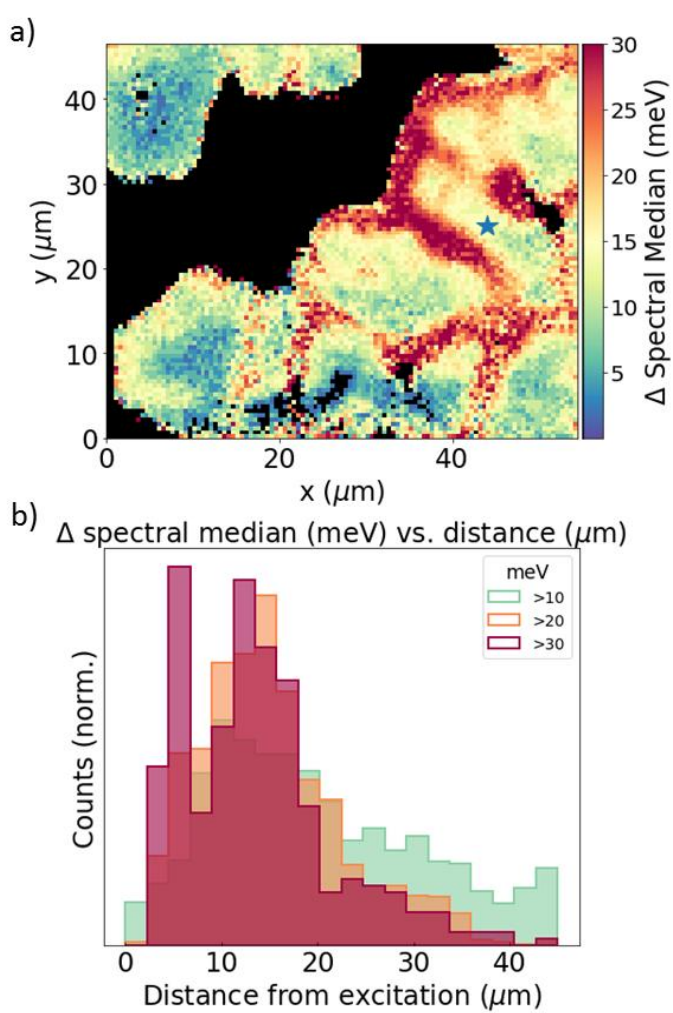

**Figure 5 - Spatial extent of photo-induced electron extraction: tens of micrometers in distance and preferentially along grain boundaries.** a) The difference map of the spectral median of a larger area of MoS$_2$ flakes illustrating variations of the spectral

median of the PL of more than 30 meV. Photodoping was performed at the blue colored star in an approximately 1 μm diameter spot. b) Normalized statistical correlation of the magnitude of the change in the spectral median (Δ) with the distance from the excitation spot. We note that the changes reach distances of around 40 μm, and that the magnitude of Δ is not correlated to the distance from the point that photodoping was conducted.

Interestingly, a correlation between the change in the spectral median and the distance from the region of the photodoping is not observed. This lack of correlation indicates that distance is not the limiting factor for the accumulation of carriers in specific locations of $MoS_2$ (see Figure 5b). Instead, there is a clear correlation between the increase in the spectral median and the initial energy of the $MoS_2$ PL with larger variations observed in localized regions with lower initial PL energy (see Fig. S7 in the Supplementary Information). As discussed above, this is indicative of the initially enhanced local carrier density prior to photodoping due to the increased contribution of trions in this region. From these observations, we hypothesize that areas with higher initial carrier densities have a local band structure that serves as a driving force for the carrier redistribution in the $MoS_2$ flakes. Notably, such areas coincide with edge regions or grain boundaries. There are several possible contributions to the edge and grain boundary-specific behavior that we observe here, including higher defect density (such as S vacancies) as well as inhomogeneous strain due to interactions with the underlying substrate.[8,9,45] Here, Raman spectroscopic imaging (Figure S 8 in the Supplementary information)[10,13,46] indicates that the edges and other regions with the largest photo-induced increase in the spectral median have lower amounts of tensile strain. Strain effects on the band structure of monolayer TMDCs are well-documented[9,13,14,46,47] showing, among others, that strain relaxation increases the quasiparticle energies for electrons and holes at the K point.[13] In our case, such an increase in the valence band energy would provide a driving force to re-equilibrate electrons from the edges and grain boundaries as electrons are extracted from the center of the flake due to the localized photodoping.[10,46,47] However, the extent of the PL changes observed by us after photo excitation are larger than the spatial variation of the strain, suggesting that the electronic structure in monolayer TMDC flakes varies more gradually towards the center of the flakes than what is observed through strain.

Finally, we give a quantitative estimate of the photo-induced carrier extraction from $MoS_2$ to the ITO nanocrystals and the processes involved. As shown in Fig. 3b, and elaborated on in Figure S 9 in the Supplementary Information, the temporal variation of the charge density strongly resembles the transient response time of a capacitor. In fact, using the initial carrier density of the $MoS_2$ from the PL analysis, and assuming that the potential difference between the ITO NCs and $MoS_2$ is commensurate with the valence band offsets, we obtain a reasonable fit to photo-induced charging of a model capacitor. We extract a capacitance value in the attofarad (*aF*) range, which is in a similar range to previous studies that have shown the charging of metal oxide nanocrystals that act as capacitors.[32,33] As shown in Figs. 4 and 5, the change in carrier density actually delocalizes over the entire flake, and favors the outer region of the flake. Analysis of the exciton/trion spectral weight over the entire area investigated in Figure 5 that takes into consideration only those areas in which the change in the spectral median is larger than 10 meV indicates that the number of transferred charges is around 40 carriers per nanocrystal. We highlight here that reaching capacitive carrier injection of 250 carriers per nanocrystal should be possible by implementing specifically designed and optimized nanocapacitors[32], and thus this effect can become even more pronounced.

## Summary

To summarize, by exploiting the ability to control free-carrier concentrations in ITO nanocrystals using only light, we drive significant carrier extraction from monolayer $MoS_2$. Optically induced carrier density changes in the range of $6\times10^{12}$ cm$^{-2}$ are achieved in the $MoS_2$, similar to gating experiments shown in literature but without the need for back contacting a device, or for bulky contacts of any kind. Notably, the effect follows a capacitor-like behavior with capacitance values in the *aF* range, leading to an average optically induced photodoping of each metal oxide nanoparticle with more than 40 carriers. This is the first demonstration of all-solid-state photodoping of such nanocrystals. Delocalized – and quasi-permanent – carrier extraction from MoS$_2$ occurs at distances up to 40 μm away from the local (micron sized) optical excitation spot, coinciding with regions of initially enhanced n-type doping and preferentially along edges and grain boundaries. The observed extent of carrier extraction from the MoS$_2$ reaches far more into the flake than do changes in local strain, pointing towards a picture where electronic structure is varying more gradually than strain. Our results suggest that control of the spatial environment serves to locally manipulate the electronic structure of TMDCs. This introduces a driving force for permanent, all optical, far reaching carrier density variations of TMDCs in hybrid structures with metal oxide nanocrystals, and thus opens new functionality and application spaces for monolayer TMDC hybrids for energy conversion, exciton funneling and quantum optics.

## Methods

### Nanocrystal synthesis and characterization

ITO/In$_2$O$_3$ core/shell nanocrystals were synthesized in a manner similar to that described in literature.[36,37] Two precursor solutions were prepared in separate vials. In one vial, tin(IV) acetate and indium(III) acetate were mixed in a 1:9 Sn:In ratio. Oleic acid was added to this vial in a 1:6 metal to acid ratio to yield a 10% Sn doped ITO precursor solution. In a separate vial, indium(III) acetate was mixed with oleic acid in a 1:6 molar ratio to yield an undoped indium oleate precursor solution. Both vials were left at 150 $^0$C under N$_2$ for several hours. The ITO nanocrystal cores were first prepared by adding the ITO precursor solution dropwise (at a rate of 0.35 mL/min) via a syringe pump to 13.0 mL of oleyl alcohol at 290 $^0$C. The ITO nanocrystal cores were grown to a size of 5.9 nm, and a small aliquot was taken out of the reaction mixture for analysis. After the addition of the ITO precursor solution, the undoped indium oleate precursor solution was added in a similar manner as described above. Enough undoped indium oleate precursor was added to allow the nanocrystals to grow to a final size of 10.1 nm. Nanocrystals were isolated by precipitating with ~12 mL ethanol. The solid was collected by centrifugation at 7000 rpm for 10 min. The solid was then collected and washed twice more with ethanol.

Nanocrystals were characterized for size and polydispersity by small-angle X-ray scattering (SAXS) and transmission electron microscopy (TEM). SAXS analysis was done on a lab-scale SAXS (SAXSess, Anton Par, Austria) equipped with an X-ray tube (Cu Kα) operating at 40 kV and 50 mA. The scattered X-ray intensities were measured with a charge-couple device detector (Roper Scientific, Germany). Raw data was processed with SAXSquant software (version 2.0). Scattering curves were averaged over 50 individual curves for various acquisition times (0.2-10s). Curve fitting was done using Irena macros for IGOR (V. 6.3.7.2).[48] TEM analysis was implemented to corroborate SAXS analysis, and was done using an FEI Tecnai Spirit TEM (Hillsboro, OR) operating at 120 kV. Nanocrystals were imaged on Ted Pella (Redding, CA) lacey carbon grids supported by a copper mesh.

### MoS$_2$ sample preparation

The MoS$_2$ monolayer film is prepared by powder vaporization technique similar to our previous work.[49] 2 mg MoO$_3$ powder (Sigma Aldirch, 99.97%) and 200 mg sulfur powder (Sigma Aldrich, 99.998%) is located in two hot zones in a home-build horizontal furnace (2'' tube diameter). Prior to the growth, the chamber is pumped to the base pressure (18mTorr) for 5 mins to remove the moisture and contaminations. During the growth, the MoO$_3$ along with the substrate is heated up to 550 °C for 2 mins for the nucleation, followed by tuning the temperature for S powder to 130 °C when the nucleation step is finished. The chamber temperature is increased to 725 °C for 15 mins while keeping the S powder at 130 °C. 100sccm Ar is employed as the carrier gas in the growth.

### Hybrid sample formation

Nanocrystal films were deposited over the MoS$_2$ sample by spin coating 20 μL of the 10mg/mL stock solution for 45 sec and 2000 rounds per minute (rpm). A thin layer of poly(methyl methacrylate) was spincoated by depositing 50 μL of a 5mg/mL stock solution for 45 sec and 2000 rpm.

### Optical measurements

PL imaging was performed on a scanning confocal microscope implementing pulsed laser excitation (λ = 500 nm; 80 MHz repetition rate) focusing onto the sample with a 50 × 0.6 NA objective to a diffraction-limited spot. Photoluminescence from the sample was collected by the same objective, and reflected laser light was filtered from the collected light using a 600 nm long-pass filter combined with a 532 nm long-pass dichroic mirror. The filtered photoluminescence dispersed by a spectrometer (Princeton Instruments) and detected with a cooled charge coupled device (CCD) camera (Andor). Mapping was performed by raster-scanning the sample with high precision piezo stages (Mad City Labs) and collecting photoluminescence spectrum at each spatial position using custom microscopy software.[50] For all excitation energies, the excitation density was kept in the range in which the PL scales linearly with excitation density. Optical photodoping was obtained by focusing a 350 nm pulsed laser onto a spot of the sample for a specific time interval. Raman characterization was performed on a Raman microscope system (NTMDT) with CW laser excitation at 532 nm and a power of 500 µW, focused by a 100× 0.6 NA objective on the sample. The emission from the sample was collected with the same objective and then analyzed using the Raman spectroscopy system.

**Fit of the temporal variation of the PL:**

The photoluminescence was fitted by two Lorentzian functions:

$$L = \frac{amp_{exc} \cdot \frac{1}{2} wid_{exc}}{\left((x-cen_{exc})^2 + \left(\frac{1}{2} wid_{exc}\right)^2\right)} + \frac{amp_{trion} \cdot \frac{1}{2} wid_{trion}}{\left((x-cen_{trion})^2 + \left(\frac{1}{2} wid_{trion}\right)^2\right)},$$

where $cen_{trion} = cen_{exc} - cen'$, and $wid_{trion} = wid_{exc} + wid'$, which ensured that the separation of the two peaks is always positive (i.e. the trion is always more red) and that the spectral widths are remaining rather similar. The initial fit was performed on the first spectrum before UV exposure defining the upper and lower limits as

$amp_{exc}$: min=0.0, max=500;

$cen_{exc}$: min=1.7, max=1.88;

$wid_{exc}$: min=0.01, max=0.1;

$amp_{trion}$: min=0.0, max=500;

cen': min=0, max=0.04;

wid': min=-0.125, max=0.125.

The temporal evolution was then fitted by keeping the widths constant and allowing only the amplitude and center wavelength change. These latter values were used as the respective initial parameters for the next fit. The constant width was a prerequisite for a reasonable comparison of intensity variations between exciton and trion over the total PL spectrum.

**Fit of the spectral variation of the PL over the entire map:**

The spectral fit over the flake was performed at each pixel in a similar way as described above. However, we were forcing the widths to remain nearly equal by setting Δwid to min=-0.0025, max=0.0025. This was necessary to obtain reasonable fits as the generally noisier data (due to the lower integration time) requested for a more stringent fit procedure. For the comparison before and after UV exposure we used the parameters extracted from the map before exposure at each pixel as the starting point for the fit of the same pixel after UV exposure, keeping the extracted width of the Lorentz peaks associated to exciton and trion constant.

**Fit of the Raman peaks:**

The Raman peaks were fitted by implementing one Lorentzian function per peak.

## Acknowledgements

This project has received funding from the European Union's Horizon 2020 research and innovation programme (MOPTOPus) under the Marie Skłodowska-Curie grant agreement No. [705444] as well as (SONAR) Grant Agreement No. (734690). Work at the Molecular Foundry was supported by the Office of Science, Office of Basic Energy Sciences, of the U.S. Department of Energy under Contract No. DE-AC02-05CH11231.

# Supporting Information

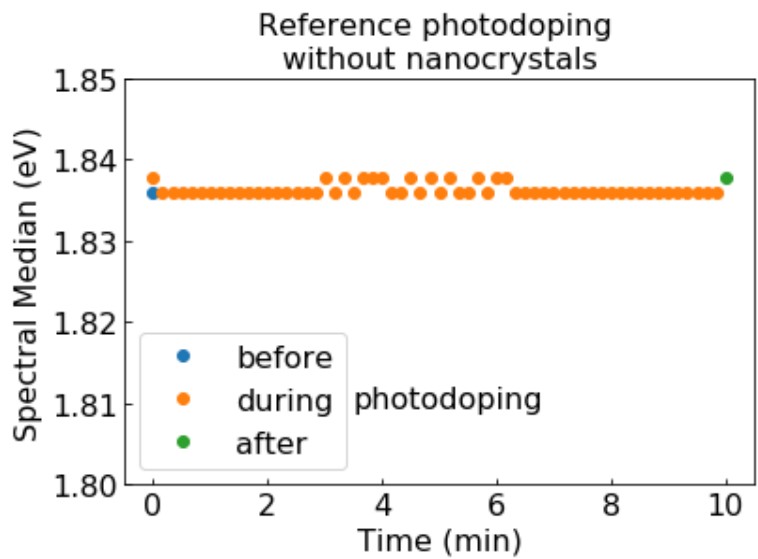

**Figure S 1 Control measurements without the ITO nanocrystals.** The PL spectrum of the monolayer $MoS_2$ does not evolve with time under the UV illumination.

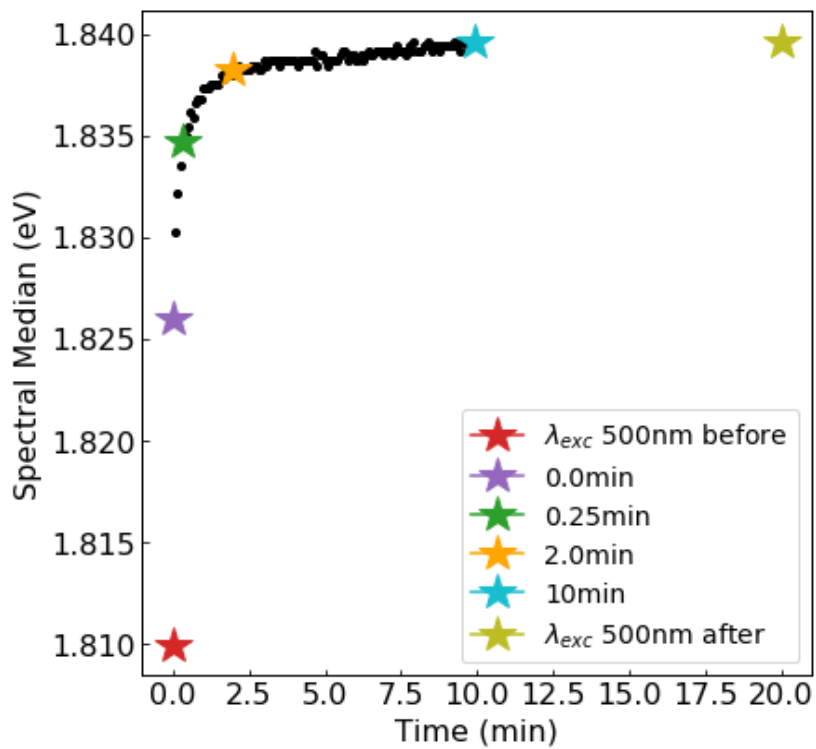

**Figure S 2 Temporal variations during p-type photodoping of $MoS_2$.** Spectral median shift of the PL peak versus exposure time. The stars in the graph correspond to the PL plots in **Error! Reference source not found.** of the Supplementary Information (similar color coding). The permanent variation of the PL was confirmed by re-measuring the $MoS_2$ PL after additional 10 minutes with 500 nm excitation (yellow star).

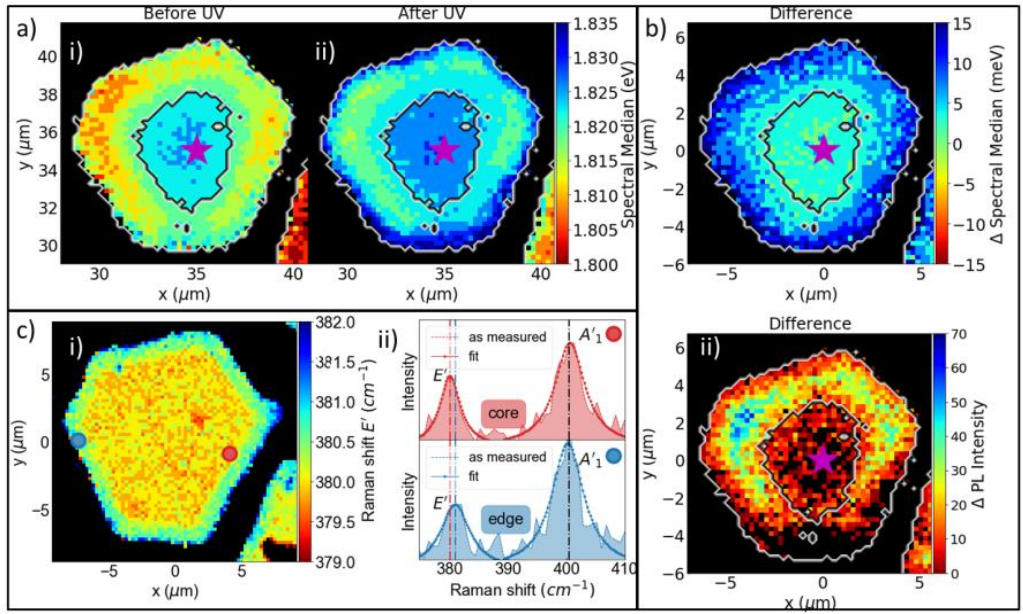

**Figure S 3. Illustration of spatial extent of spectral median variation upon photodoping in another typical example.** a) Illustration of the overall variation of the PL over an MoS$_2$ flake a i) before and a ii) after exposure to a focused UV laser for 5 minutes at the location marked with a star (laser spot size of approximately 1x1 μm in diameter). Essential differences in the spectral median between the edge and core region of the flake (illustrated by black separating lines in the plots) are obvious already before photodoping and major variations after photodoping are observed only in the edge region. b i) Plot of the difference of the spectral median of the PL after exposure to UV light illustrating a blue shift after photodoping reaching values around 15 meV, particularly obvious in the edge region of the flake, while the core region remains fairly unchanged. ii) It is notable that also the intensity of the overall PL in the edge region increases significantly after UV exposure with respect to the core region as illustrated in the difference map of the PL intensity. c) Raman shift of the E' mode over the same MoS$_2$ flake: higher cm$^{-1}$ values in the edge region are observed, corresponding to less tensile strain, correlating with the maximum variation of the spectral median in the flake after UV exposure. ii) Example of two Raman spectra in the core (red) and edge (blue) region, illustrating that mostly the E' mode is affected by strain.

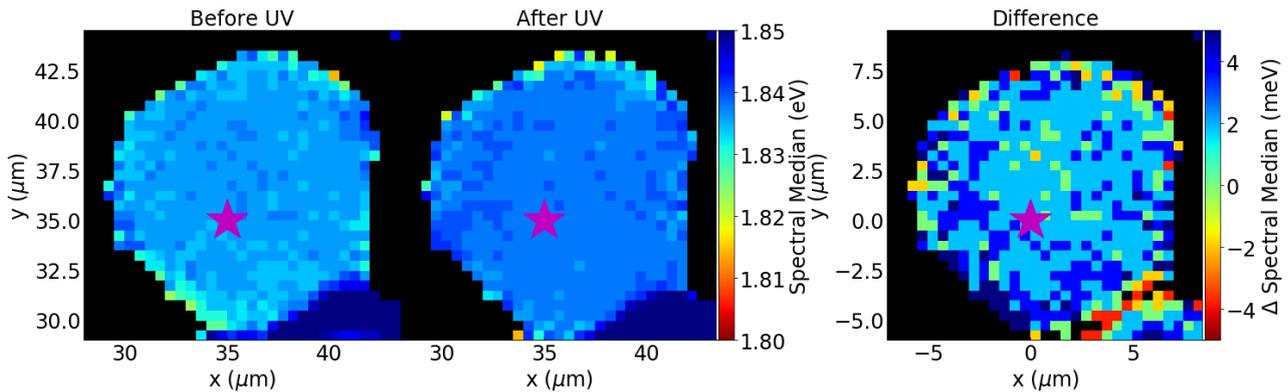

**Figure S 4 Photodoping of a reference sample without ITO NCs.** Plot of the spectral median of the PL spectrum scanned over a single MoS$_2$ flake without ITO NCs before and after exposure to UV light. No significant variations are observed as demonstrated by the difference map. The pink star illustrates the location at which the UV (350 nm) laser was focused (around 1x1 μm in diameter).

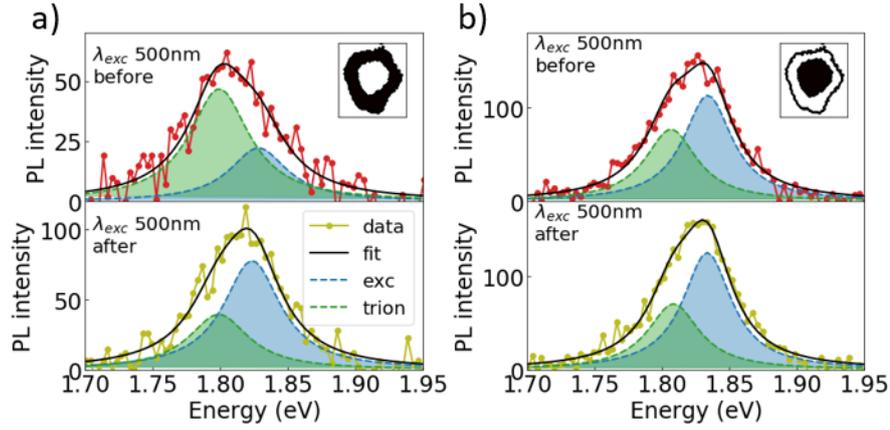

**Figure S 5 Representative spectra before and after p-type photodoping in the core and edge region of the flake.** Typical spectra before (red) and after (yellow) photodoping together with their respective fits (black curves) for the edge (a) and core (b) region. A strong increase of the spectral weight of the exciton peak (blue shaded curve) at the expense of the trion peak (green shaded curve) is observed in the edge region, while in the core region no significant changes are identified.

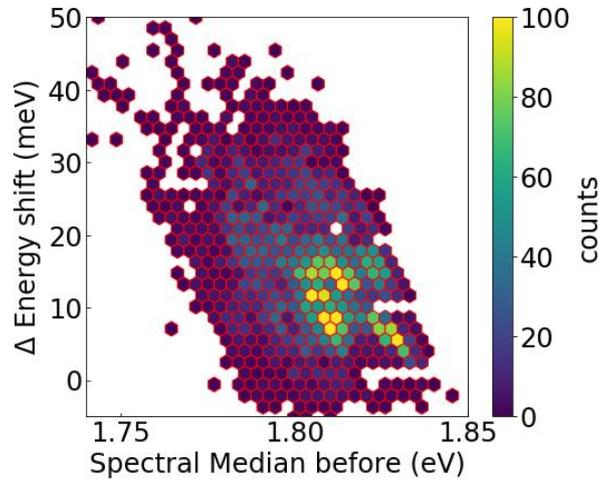

**Figure S 6** Correlation between the initial spectral position of the PL peak (eV) and the PL shift after photodoping (Δ spectral median, meV). The highest variations correlate to an initially lower energy PL. Brighter colors indicate higher counts for the specific correlation.

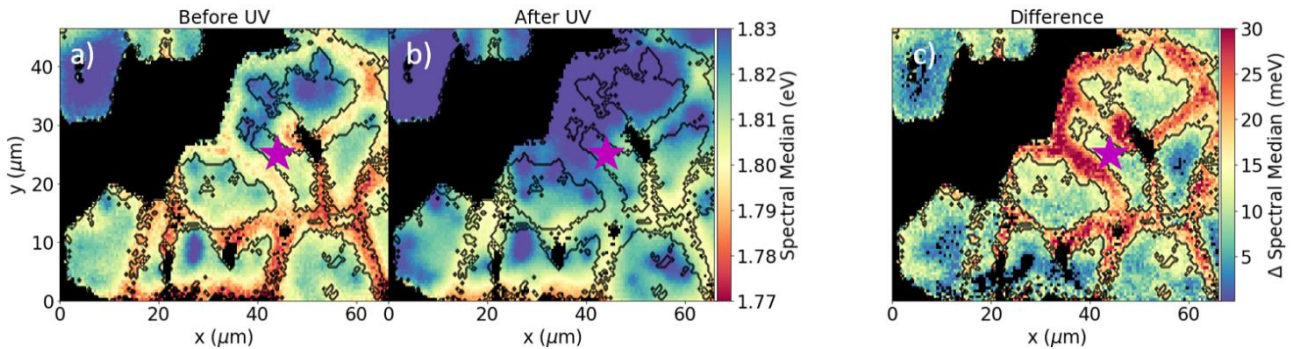

**Figure S 7 Spectral map of larger area of monolayer MoS2 flakes merged to each other.** Variations in spectral median a) before, b) after photodoping and c) the difference map.

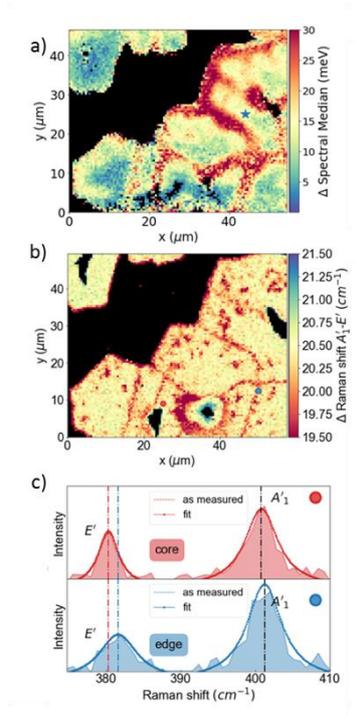

**Figure S 8 Raman measurements illustrating relaxed strain along grain boundaries.** A measure of strain in monolayer $MoS_2$ is the Raman shift of the E' and $A_1$'. Lower Δ Raman shifts of E' and $A_1$' indicate relaxed tensile strain. In fact, in our Raman map of the same area of investigation grain boundaries of the individual $MoS_2$ flakes are highlighted by lower Δ Raman shifts, i.e. relaxed tensile strain, and coincide with the largest photo-induced electron extraction. a) The difference map of the spectral median of a larger area of $MoS_2$ flakes illustrating variations of the spectral median of the PL of more than 30 meV. Photodoping was induced at the blue colored star in a diffraction limited spot. b) Δ Raman map of the difference of $A_1$' and E' modes showing lowest Δ Raman shift along the grain boundaries, correlated to regions with decreased strain. c) Example of two Raman spectra in the core (red) and edge (blue) region, illustrating that only the E' mode is affected by strain. Spectra correspond to the red and blue dots in Figure 4c.

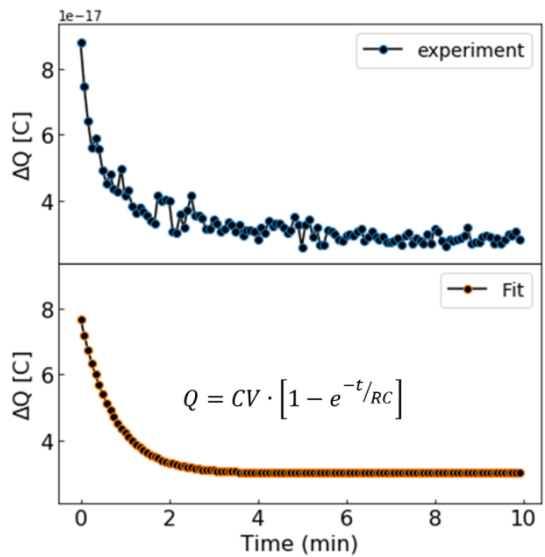

**Figure S 9 Charge density variation ΔQ.** Charge density variation extracted from the analysis of exciton/trion PL ratio together with a fit.